\documentclass[12pt]{article}
\usepackage[utf8]{inputenc}
\usepackage{amsmath, amssymb}
\usepackage[sorting=none, bibstyle=numeric, backend=bibtex]{biblatex}
\bibliography{my.bib}
\usepackage[english]{babel}

\usepackage{graphicx, calc}
\usepackage{subfig}
\usepackage{float}
\usepackage{authblk}

\title{Exact loop densities in   $O(1)$ dense loop model on the cylinder of odd circumference and clusters in half-turn self-dual critical percolation}

\author[1,2]{A.M. Povolotsky\thanks{alexander.povolotsky@gmail.com}}
\author[3]{A.A. Trofimova \thanks{anastasiia.trofimova@gssi.it, nasta.trofimova@gmail.com}}

\affil[1]{Bololiubov Laboratory of Theoretical Physics, Joint Institute for Nuclear Research, 141980, Dubna, Russia }
\affil[2]{National Research University Higher School of Economics, 20 Myasnitskaya, 101000, Moscow, Russia }
\affil[3]{Gran Sasso Science Institute Viale Francesco Crispi 7, 67100, L’Aquila, Italy}

\date{}

\newcommand*\inlinegraphicsformula[1]{%
	\settototalheight\myheight{Xygp}%
	\settodepth\mydepth{Xygp}%
	\raisebox{-2\mydepth}{\includegraphics[height=1.1\myheight]{#1}}%
}

\newlength\myheight
\newlength\mydepth
\settototalheight\myheight{Xygp}
\newcommand*\inlinegraphics[1]{%
	\settototalheight\myheight{Xygp}%
	\settodepth\mydepth{Xygp}%
	\raisebox{-3.5\mydepth}{\includegraphics[height=2.5\myheight]{#1}}%
}

\begin{document}
\emergencystretch 3em
	
	\maketitle
	\begin{abstract}
		We consider   $O(1)$ dense loop model in a square lattice wrapped on a cylinder of odd circumference $L$ and calculate the exact densities of loops. These densities of loops are equal to the densities of critical bond percolation clusters on a forty-five-degree rotated square lattice rolled into a cylinder with special boundary conditions which we refer to as half-turn self-dual percolation. 
		The solution is based on a correspondence between the   $O(1)$ dense loop model and the six-vertex model at the Razumov-Stroganov point. 
	\end{abstract}
	
	\section{Introduction}
	Percolation theory is a thriving field at the intersection of statistical mechanics and probability theory studying one of the simplest though extremely rich examples of phase transitions and thus serving as a testing lab for the theory of critical phenomena.     
	The bond percolation model is defined in terms of a graph in which bonds are independently chosen to be open with probability $p$ or closed with probability $1-p$. Open bonds form connected clusters,  whose statistics is the main subject of studies. In particular, when the underlying graph is infinite,  a typical situation is that a giant connected cluster appears when $p$ exceeds a certain critical value  $p_c$.
	This phenomenon is coined  percolation transition. 
	
	As usual in the theory of phase transitions and critical phenomena, one is interested in the statistics of physical observables, some of which are expected to reveal a universal behaviour in the vicinity and at the critical point.  In particular, significant progress was made in studies of the percolation transition at two-dimensional periodic planar lattices. The statistics of con\-nec\-ted clusters in this case is expected to possess conformal invariance,  which severely restricts the range of critical indices and scaling functions characterizing their large-scale behavior. Though in a few cases the conformal invariance of planar percolation was proved mathematically rigorously,
	exact solutions still remain the major tool for testing the predictions of conformal field theory. In particular, the critical bond percolation models on square and triangular lattices have the advantage of being exactly solvable due to their relation to the exactly solvable six-vertex model.
	There are also particular cases of the Fortuin-Kasteleyn random cluster model in turn related to the Potts model, which is also solvable at the critical point due to the same connection.   
	
	A particular example of the quantity of interest, which we focus on in the present paper, is the average number of connected clusters per lattice site.    This quantity has been widely studied since long ago, first approximately at arbitrary $p$,  using the series expansions  \cite{sykes1963some,sykes1964exact}, and then at the critical point, $p=p_c$,  using the exact solutions \cite{baxter2016exactly}.
	In particular exact infinite plane limits of these densities were obtained in \cite{lieb19711,baxter1978triangular}. 
	
	Though these limits are not universal, e.g. lattice dependent, the universality is expected to be manifested in the finite-size corrections to the bulk values of the cluster densities, when the model is considered in confined geometry. The conformal field theory (CFT) predicts universal finite-size corrections, which depend on the boundary conditions (BCs). Their values were conjectured  within  the Coulomb gas theory \cite{Nienhuis1987}  and can be found using the exact Bethe ansatz solution of related models \cite{ABB, HQB}.  In particular, the CFT-based leading finite-size corrections to critical percolation cluster densities on the infinite strip with free and periodic BCs were conjectured and numerically checked in  \cite{ZiffFinchAdamchik, KlebanZiff}. 	
	
	The exact values of the densities of connected clusters in confined geometry, which in particular would allow one to test the predicted finite-size corrections, stayed unknown until recently. One approach to studying their dependence on the open bond probability $p$ was undertaken in \cite{ChangShrock2004, ChangShrock2021}.
	However, being based on manipulations with a finite dimensional transfer matrix it allowed obtaining exact cluster densities for only a few small values of the strip width. No closed formulas for arbitrary lattice sizes were obtained in this way, though the densities were proved to be rational functions of $p$.
	
	In the present paper, we employ another approach suitable for studying bond percolation at the square lattice exactly at the critical point $p=p_c=1/2$. It is based on the tight connection of the critical percolation  with the  $O(1)$ dense loop model (DLM)
	\cite{Zinn2009six} that in turn is related to the six-vertex model \cite{Lieb1} at specific values of parameters coined combinatorial or stochastic point.
	The transfer matrices of the models at this point can be associated with a Markov chain whose stationary state has a remarkable structure discovered by Razumov and Stroganov \cite{RS-01, RS-04}. They observed that the stationary state probabilities can be rescaled to become integers having a purely combinatorial meaning. These observations 
	revealed a possibility of obtaining  exact finite-size formulas for the ground state observables of the model and 
	triggered a burst of activity in establishing  connections between the percolation, the   $O(1)$ DLM, the six-vertex
	model, the XXZ model, the fully packed loop model, and alternating
	sign matrices \cite{BGM,RS-04,RS-05,G}. These studies   resulted in  plenty of conjectures
	and exact results  for finite lattices with various BCs \cite{FZ2005,FZ2005_2,Z2006,FZZ2006,FZ2007,RSZ2007,CS,GBNM,MNGB-1,GierJacobsenPonsaing,MitraNienhuis}.
	In particular,  the findings by  Razumov and Stroganov suggest that the averages of integer-valued observables over the stationary state should be given by rational numbers. 
	
	Exact rational values of the average densities of contractible and non-con\-trac\-tible loops in  $O(1)$  DLM on the square lattice wrapped into an infinite cylinder of arbitrary circumference were recently obtained in \cite{Pov2021}. The configurations of this model, consisting of paths passing through every bond of the square lattice and making a ninety-degree turn at every site, 
	are in one-to-one correspondence with configurations of percolation clusters on another square lattice of forty-five-degree rotated orientation also wrapped into a cylinder, and the uniform measure on path configurations corresponds to the critical $p=1/2$ point of the percolation model. Then,  the densities of loops formed by paths can be interpreted as the densities of critical percolation clusters that do or do not wrap around the cylinder.  The asymptotic expansion of the obtained exact rational values at large cylinder circumference indeed reproduced the irrational limiting value obtained in \cite{lieb19711}   and the leading finite-size corrections predicted in \cite{KlebanZiff} for total density of all loops using the Coulomb gas theory \cite{Nienhuis1987} and for non-contractible loops \cite{alcaraz2014noncontractible} using the Bethe ansatz results for the related XXZ quantum spin chain \cite{ABB, HQB}, which also fit into the CFT framework \cite{destri1989twisted}.   These results were extended to the square lattice with arbitrary tilt \cite{povolotsky2023exact} wrapped into a cylinder with helicoidal BCs. This in particular allowed the calculation of critical percolation cluster densities on the lattice of standard orientation considered by other authors \cite{ChangShrock2004, ChangShrock2021} for small circumferences. Though  \cite{povolotsky2023exact}  did not provide a closed formula for the densities, which would be suitable for asymptotic analysis, it still allowed systematic high-precision numerical studies of the dependence of leading universal and the sub-leading non-universal finite-size corrections on the tilt. 
	
	The two last-mentioned results for the densities of loops of  $O(1)$ DLM are however limited to strips of even width with periodic and helicoidal BCs, i.e. the cylinders with even circumferences only. It is also the case, in which the  $O(1)$ DLM path configurations can be mapped to the percolation on the rotated lattice with periodic or helicoidal BCs. The situation turns out to be very different when one considers DLM on the cylinder of odd circumference. 
	
	A significant distinction between the odd and even number of sites in the spin chains has been observed by several authors  \cite{faddeev1981spin,baake1988higher, Str2001importance, RStr2001, PStr1999}. In particular, the structure of the spectrum of the Hamiltonian depends drastically on whether the chain has an even or odd number of spins. The discrepancy also applies to the related lattice models. In particular, it can be seen in the structure of  $O(1)$ DLM path configurations:  on the infinitely long cylinder of even circumference the configurations almost surely consist only of finite loops,  while on the cylinder of odd circumference at least one infinite path called defect exists preventing non-contractible loops from closing around the cylinder.  Therefore, though the infinite plane limit of the cluster densities in odd and even cases coincide, already the leading finite-size corrections, expected to be universal CFT-determined quantities depending on the BCs, are distinct.
	It is indeed the case for the finite-size corrections to similar quantities obtained for the continuous time relatives
	of the mentioned lattice models, XXZ model \cite{ABB, HQB, Str2001importance} and Raise and Peel stochastic process  \cite{povolotsky2018large,povolotsky2019laws}. In addition, the mapping of  $O(1)$ DLM  on the cylinder of odd circumference to the periodic critical percolation breaks up and needs a modification of  BCs.  Therefore, the 
	odd case of  $O(1)$ DLM as well as the related percolation model, if it can be defined, requires a separate consideration.

	In the present paper, we extend the results of  \cite{Pov2021} to the  $O(1)$ DLM to  the  case of a cylinder with an odd circumference 
	obtaining the exact loop density as a function of the circumference. We also establish the mapping of path configurations in this case to the percolation on the cylinder with special BCs, which we refer to as the half-turn self-dual percolation.
	As before, the loop density in  $O(1)$ DLM becomes the density of percolation clusters at the critical point under this mapping.   Using the formulas obtained we derive a few first terms of the asymptotic expansion of the densities. In particular, we reproduce the infinite plane density limit and obtain the leading finite-size correction to it, which is compared to the even case and the continuous time analogs. The solution is based on the mapping of  $O(1)$ DLM to the six-vertex model. It follows the line of \cite{FSZ2000even, FSZ2001even} and especially of  \cite{Str2001importance}, where specific solutions of Baxter's T-Q equation \cite{baxter1972partition} for the ground states of the    XXZ model with $\Delta=\pm1/2$  for the chains with even and odd number of spins respectively were studied. 
	
	The article is organized as follows. Section 1  introduces the   $O(1)$ DLM on a cylinder of odd circumference and explains its relation to the critical percolation model. We present the obtained exact formulas for the loop densities and compare the case of an odd circumference with an even one. Then, in section 2 we exploit the connection between   $O(1)$ DLM and the six-vertex model at the level of transfer matrices and apply algebraic Bethe ansatz to obtain exact formulas for loop densities.    
	\newpage
	
	\section{O(1) DLM and percolation}
\begin{figure}[H] 
		\centering
		\inlinegraphics{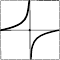}
		\hspace{0.1\textwidth}		
		\inlinegraphics{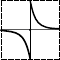}
     \caption{Two local vertex configurations in DLM.}
     \label{fig: loop-vertices}
	\end{figure}

\begin{figure}[h]
		\centering
		\includegraphics[width=0.45\textwidth]{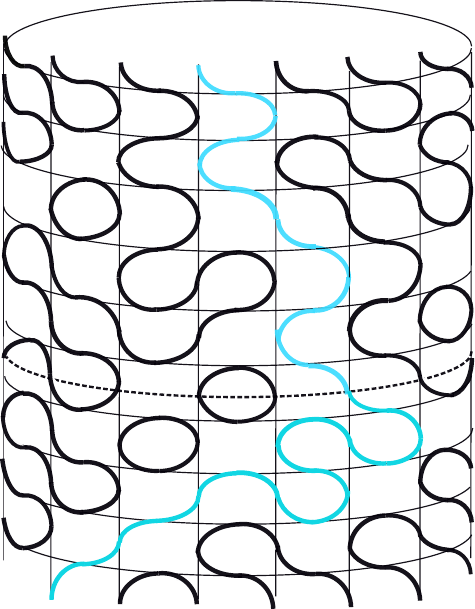}
		\caption{A typical loop configuration on an infinite cylinder of odd circumference $L = 7$. The defect line of the lower semi-infinite cylinder is colored blue. The vertices on its boundary are labelled from left to right. The chord diagram shows how the vertices are connected to each other.}
		\label{fig: cylinder}
	\end{figure}

	We define the $O(n)$ DLM on a strip $\mathbb{Z}/L\mathbb{Z} \times \mathbb{Z}$ of a two-dimensional square lattice wrapped into an infinite cylinder, i.e. with periodic boundary conditions in the horizontal direction. Here, we assume the circumference $L$ to be odd $L=2N+1, N\in\mathbb{N}_0$. Configurations of the model consist of  lattice paths passing along  every bond of the lattice and making a ninety-degree turn at every site.  
	The path configurations are supposed to be distributed according to a measure that assigns  (unnormalized) Boltzmann  weight $n$ to every closed loop and a unit weight  to  infinite paths, if there are any. Here we focus on the  $O(1)$ DLM,   which means that   the measure on path configurations is uniform.   In this case, one  can also construct a random path configuration by solely local operations placing   one of two  vertices with a pair of arcs shown in Fig.\ref{fig: loop-vertices},  chosen  with  equal  probabilities  independently at every site of the lattice.
	It is obvious that while under this weighting  only finite loops exist with probability one  on the infinite cylinder with even circumference, there is exactly one  infinite path at the odd one.  
	To see  this, consider the   infinite in the vertical direction cylinder and cut the vertical bonds   between any two subsequent   horizontal site rows. A random lattice path composed of  arcs   starting  from a cut bond    and  walking in the upper half of the cylinder  returns  to another  cut bond  with probability one. So do  paths in the lower half of the cylinder. Therefore, vertical links at any horizontal level are paired by half-loops in the upper and lower half-cylinders, provided there is a match for each of them, which is the case when the number of cut bonds is even.   This guarantees that any path is a loop with probability one. On the other hand, when
	the number of vertical bonds at a horizontal cut is odd, there is no  match for  one of them, and hence the corresponding path  should go upward and downward to infinity. Of course, one can draw more than one path going to infinity. However, due to  the random walk  recurrence in one  and two dimensions,  such configurations on an infinite cylinder have zero measure.   
	The presence of such an infinite path, which we refer to as a defect, does not allow any  loop to wind around the cylinder, making all loops contractible 
	(see Fig. \ref{fig: cylinder}).
	
	\begin{figure}[h]
		\centering
		\includegraphics[width=0.8\textwidth]{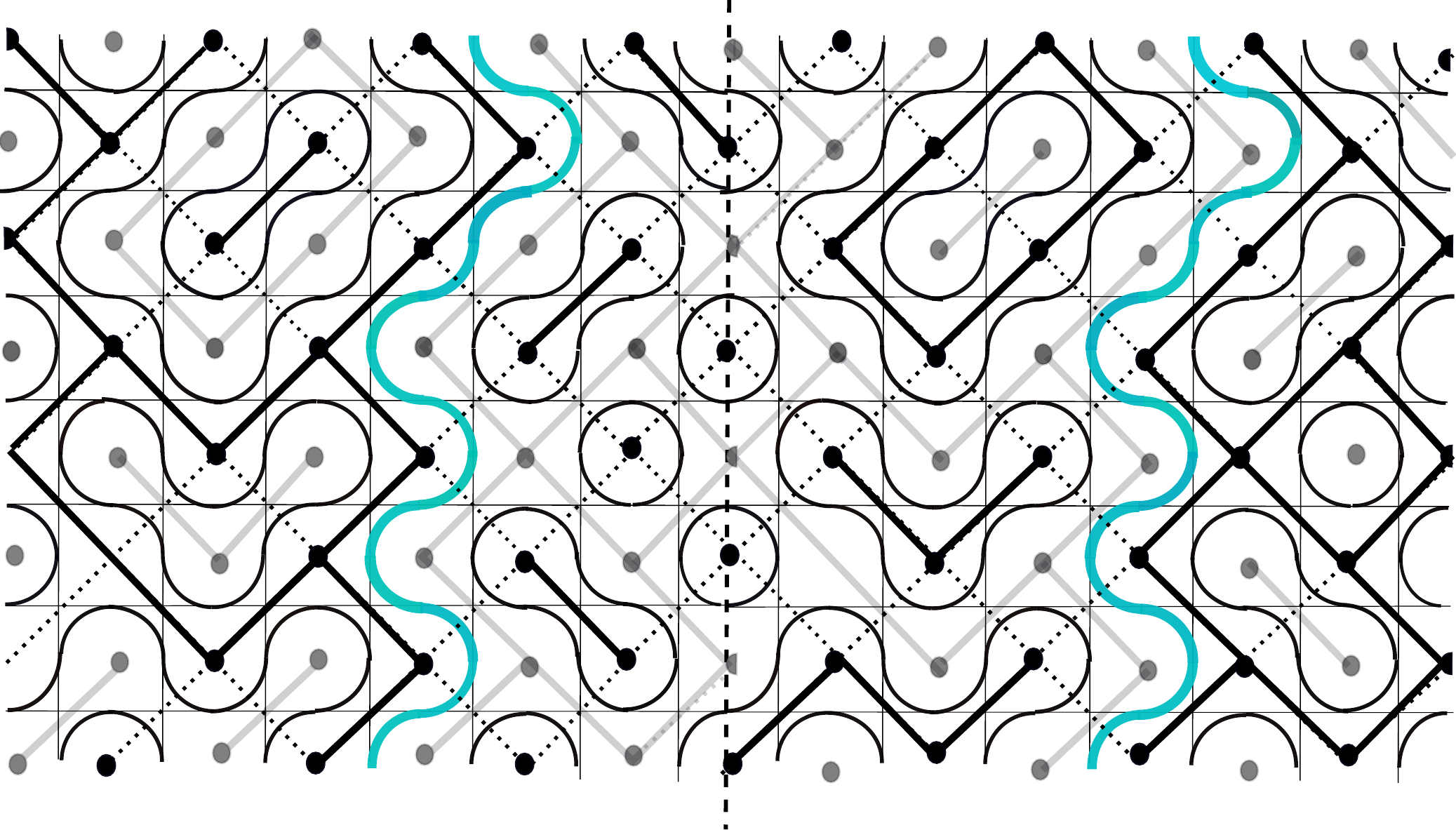}
		\caption{Correspondence between the configuration of the   $O(1)$ loop model (pictured twice side-by-side in thin black solid lines with the defect line shown in blue) and the associated  percolation cluster (thick solid black lines) on the rotated lattice.  There is a dual percolation cluster pictured in thick grey solid lines.}
		\label{fig: correspondence DLM and percolation}
	\end{figure}
	
	The mapping between   $O(1)$ DLM and critical percolation clusters can be constructed in the following way. 
	The percolation is considered on a forty-five-degree rotated square lattice, for which the original lattice is the so-called medial graph. This means that every site of the original lattice is placed in the middle of the bond  of the rotated lattice, while 
	the  sites of the rotated lattice are at  the center of  every second face of the original lattice arranged  in checkerboard order.  
	The open bonds   of the percolation configuration  are then between two arcs of  DLM and  the bonds traversed by a pair of arcs are closed. 
	
	The above description does not yet take the boundary conditions into account. If one assumes the periodic boundary conditions associated with the DLM on  the cylinder of even circumference, the corresponding rotated lattice is also rolled into a cylinder, see e.g. \cite{Pov2021}. However, it is not the case for an odd circumference. Can we still map it to any version of percolation? The route around this problem is to duplicate 
	the original DLM lattice, placing its two copies next to each other in the horizontal direction, and to roll the  doubled lattice into the cylinder as shown in Fig. \ref{fig: correspondence DLM and percolation}. Then, the  doubled DLM configuration  is  mapped as explained above to the percolation on the forty-five-degree rotated lattice also rolled into the cylinder. Of course, produced from two copies of one DLM configuration, the  percolation configuration  is not  arbitrary any more. Instead,  it possesses a special symmetry  turning into the dual one under the half rotation of the cylinder that exchanges the sites and the faces of the rotated lattice. Thus, we call percolation on the cylinder with so defined  BCs the half-turn self-dual percolation.  One can see in Fig. \ref{fig: correspondence DLM and percolation} that  configurations of  the half-turn self-dual percolation  always contain one or two  infinite connected clusters. The two infinite  clusters  include  infinite chains of  open bonds of the rotated  lattice adjacent  to two copies of the infinite defect in the DLM configurations, one  to the right of one copy and the other  to the left of the other. If these two chains are connected with each other, there is only one infinite cluster.

	Finally, we note that the  density of loops and of finite percolation clusters obtained under the mapping described are equal. Indeed, every loop is either circumscribed on a percolation cluster or is inscribed into a circuit inside a percolation cluster and hence is circumscribed on the dual cluster.  The doubling of the DLM configuration simply doubles the number of loops as well as  sites of the DLM  lattice. Since at the critical point $p=p_c=1/2$ the average number of clusters and dual clusters coincide, while the number of sites of the rotated lattice is twice less than that of the original lattice, the density of loops in  $O(1)$ DLM on the original odd circumference cylinder is  equal to the density of connected percolation clusters  on the rotated lattice obtained from the doubled DLM configuration.
	
	\subsection*{Results and discussion}
	Here we present and discuss the  main result of the article,  the density $\nu(L)$ of  loops in  $O(1)$ DLM on a strip of odd width of the square lattice with periodic boundary conditions, i.e. rolled  into a cylinder of odd circumference $L=2N+1$.  It is also   the density of  the half-turn self-dual percolation clusters in   the forty-five-degree rotated lattice    rolled into a cylinder with as many  sites in a row. The density is given by the following formula
	\begin{eqnarray} \label{eq: density_result_in_Gamma}
		\nu (2N + 1) = \frac{1}{1+2N}\left( \frac{\Gamma(\frac{N}{2})\Gamma(\frac{3}{2}+\frac{3N}{2})}{\Gamma(\frac{3N}{2})\Gamma(\frac{1}{2}+\frac{N}{2})}+
		\frac{\Gamma(\frac{1}{2}+\frac{N}{2})\Gamma(2+\frac{3N}{2})}{\Gamma(1+\frac{N}{2})\Gamma(\frac{1}{2}+\frac{3N}{2})} \right) - \frac{5}{2} \\ \nonumber
		= \frac{1}{12}, \frac{37}{400}, \frac{597}{6272}, \frac{2441}{25344}, \frac{78035}{805376}, \dots,
	\end{eqnarray}
	where $N=1,2,\dots$, and  we show its rational numerical values of $\nu (2N + 1) $ for $N = 1, 2, 3, 4, 5$. Also  it is obvious that  $\nu(1)=0$, since no loops exist on the lattice with a single vertical line, while only a single infinite defect remains. 
	This corresponds to the only infinite chain of open bonds  in the corresponding  half-turn self-dual percolation configuration. Though the same  result,   can formally be obtained as the $N\to0$  limit of the formula (\ref{eq: density_result_in_Gamma}), it should be tackled separately,  since the formula is applicable only for $N\geq 1$.  
	
	Similarly to the even case  the densities are rational numbers, which can be seen from yet another form of (\ref{eq: density_result_in_Gamma}) written in terms of Pochhammer symbols $(a)_n = a (a+1)\dots (a+n-1),$ 
	\begin{equation}
		\nu (2N + 1) = \frac{1}{1+2N}\left( \frac{(\frac{1}{2}+\frac{N}{2})_{N+1}}{(\frac{N}{2})_{N}}+
		\frac{(1+\frac{N}{2})_{N+1}}{(\frac{1}{2}+\frac{N}{2})_{N}} \right) - \frac{5}{2}, \,\, N = 1, 2,  \dots  
	\end{equation}
	
	The representation \eqref{eq: density_result_in_Gamma} is suitable for an asymptotic expansion where the application of Stirling formula to the gamma functions yields
	\begin{equation} \label{eq: density_odd_serie}
		\nu (L) = \frac{3\sqrt{3}-5}{2} - \frac{1}{4 \sqrt{3}} L^{-2} + \frac{35}{144 \sqrt{3}} L^{-4} + O(L^{-6}).
	\end{equation}
	As expected, the infinite plane limit $\nu(\infty)=(3\sqrt{3}-5)/2$  is as obtained in  \cite{lieb19711,baxter1978triangular}. The $O(L^{-2})$ finite-size correction is expected to be  universal being the CFT-prescribed conformal anomaly of the theory.
	
	It is informative to compare the result with the density of contractible loops on the cylinder of even circumference $L=2N$, found in \cite{Pov2021}, 
	which can be written in a form similar to  \eqref{eq: density_result_in_Gamma}
	\begin{equation}
		\nu_c (2N ) =  \frac{3 \Gamma(\frac{N}{2}) \Gamma(\frac{1}{2}+\frac{3N}{2})}{4\Gamma(\frac{3N}{2})\Gamma(\frac{1}{2}+\frac{N}{2})}+
		\frac{9 \Gamma(\frac{1}{2}+\frac{N}{2})\Gamma(\frac{3N}{2})}{4\Gamma(\frac{N}{2})\Gamma(\frac{1}{2}+\frac{3N}{2})}-\frac{5}{2}.
	\end{equation}
	Its asymptotic expansion 
	\begin{equation} \label{eq: density_even_serie}
		\nu_c (L) = \frac{3\sqrt{3}-5}{2} + \frac{1}{4 \sqrt{3}} L^{-2} - \frac{23}{48 \sqrt{3}} L^{-4} + O(L^{-6})
	\end{equation}
	contains the  $O(L^{-2})$ finite-size correction, which coincides with that in formula  \eqref{eq: density_result_in_Gamma}  in absolute value but has the  opposite sign. This is consistent  with the results for the Hamiltonian aka continuous-time version of DLM  obtained  in \cite{Str2001importance,MNGB-1,povolotsky2018large,povolotsky2019laws} in context of XXZ model, continuous time DLM and Raise and Peel model. 
	Note that  these continuous time limits  contain   arbitrary dimensional  factors responsible for the  continuous time scale. As a result, the CFT-based finite-size correction to the dominant eigenvalue of the Hamiltonian or Markov generator  differs from the similar correction to  the eigenvalue of the related transfer matrix by a factor sometimes referred to as the  speed of sound. This factor, though non-universal, depends only on the time scaling rather than  on the boundary conditions \cite{von1987ashkin}.    It follows that the ratio between the corrections in two models with different boundary conditions should be the same for both the Hamiltonian and for the lattice model. 
	Indeed, the leading finite-size corrections to the analogs of the  contractible loop densities, which are the correlation functions of projections of  the nearest neighboring spins on the direction transverse to the z-axis for XXZ model \cite{Str2001importance}, 
	minimal  one nest probability in the connectivity structure of  $O(1)$  DLM model \cite{MNGB-1} and the local avalanche current in the Raise and Peel model \cite{povolotsky2018large,povolotsky2019laws}, have the same absolute values and the opposite signs in the odd and even circumference cases. 
	
	The next corrections are not universal and, as one can  see,  are different between  the even and odd cases. They still are expected to have a CFT meaning  capturing  the effects  
	of breaking  the rotational symmetry of the theory by the  lattice, which should lead to the appearance   of  operators, whose  conformal spin is a multiple of four, see  \cite{Cardy1988}. The presence of these operators is revealed in the model on the lattice with an arbitrary tilt by the dependence of the finite-size corrections on the tilt angle via trigonometric functions of an argument
	that is a multiple of quadruple tilt angle,  \cite{CJV2017,TDJ2019,Pov2021}. Study of this dependence requires  consideration of the  $O(1)$ DLM on a tilted lattice rolled into a cylinder of odd circumference, as it was done in \cite{Pov2021} for the even one, and constructing an asymptotic expansion  for the tilt dependent loop/cluster densities, which is the matter of future investigation.

	\section{From  O(1) DLM to six-vertex model }
	
	In this section, we explain the relation between   $O(1)$ DLM and the six-vertex model. It allows us to express   the average loop density $\nu(L)$ in terms of the largest eigenvalue of the transfer matrix  of the associated six-vertex model.
	
\begin{figure}[H] 
		\begin{minipage}{1\linewidth}
			\centering
			$\begin{array}{c @{{} \qquad {}} c @{{} \qquad {}} c @{{} \qquad {}} c@{{} \qquad {}} c@{{} \qquad {}} c}
				\vspace{0.5cm}
				\inlinegraphics{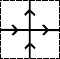} &  \inlinegraphics{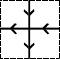} &  \inlinegraphics{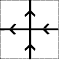} &
				\inlinegraphics{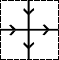} &  \inlinegraphics{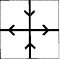} &  \inlinegraphics{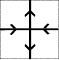}  \\[\jot]
				\vspace{0.5cm}
				a_1 & a_2 & b_1 & b_2 & c_1 & c_2  
			\end{array}$
		\end{minipage}
		\caption{The six types of vertices in the six-vertex model and their corresponding weights.}
		\label{fig: 6V weights}
	\end{figure}

	The six-vertex model is a  vertex model on the square lattice. The states of the model are defined by giving an orientation   to every bond of the lattice, i.e.  by drawing   arrows at the bonds in all possible ways subject to the so-called ice rule:  the numbers of arrows incoming to and outgoing from every site  are equal.  Thus, the states can be constructed by placing  one of six possible vertices  shown in Fig. \ref{fig: 6V weights} at every lattice site so that the direction of arrows at two sites connected by a bond agree.
	
	Each of the six vertices is assigned its own Boltzmann weight, so that the weight of a state $C$ of the model at a finite lattice domain $D$ is given by the product of   weights $w_i \in \{a_1, a_2, b_1, b_2, c_1, c_2\}$ of individual  vertices within this domain
	\begin{equation}
		W(C)=\prod_{i\in D} w_i,
	\end{equation}
	while the probability assigned to $C$ is 
	\begin{equation}
		\mathbb{P}(C)=\frac{W(C)}{Z},
	\end{equation}
	where the normalization factor 
	\begin{equation}
		Z=\sum_{C\in \Omega(D)} W(C)
	\end{equation}
	referred to as  partition function is  the sum of weights over  the set $\Omega(D)$ of states.

	To establish a connection between the six vertex model and  $O(n)$ DLM, we consider a directed  DLM,
	with  path  configurations obtained from those of  $O(n)$ DLM  by  giving one of two possible orientations  to every path. If we assign the weight of $q$ to every contractible loop oriented clockwise and the weigh  $q^{-1}$ to the one oriented anti-clockwise, the summation over loop orientations will yield the weight
	\begin{equation} \label{eq: weight n}
		n = q+ q^{-1},
	\end{equation} 
	of the undirected loop. In particular at the stochastic point 
	\begin{equation} \label{stoch_point}
		q=e^{\frac{\mathrm{i}\pi} {3}}.
	\end{equation}
	we have the unit loop weight of  $O(1)$ DLM. 
	
\begin{figure}[H] 
		\begin{minipage}{1\linewidth}
			\centering
			$\begin{array}{c @{{} \qquad {}} c @{{} \qquad {}} c @{{} \qquad {}} c@{{} \qquad {}} c@{{} \qquad {}} c@{{} \qquad {}} c@{{} \qquad {}} c}
				\vspace{0.5cm}
				\inlinegraphics{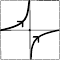} &  \inlinegraphics{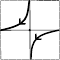} &  \inlinegraphics{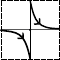} &
				\inlinegraphics{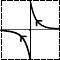} &  \inlinegraphics{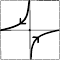} &  \inlinegraphics{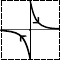} &  \inlinegraphics{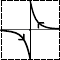} &  \inlinegraphics{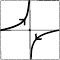} \\
				z & z& 1&1 &zq^{\frac{1}{2}} &q^{-\frac{1}{2}} & q^{\frac{1}{2}}&zq^{-\frac{1}{2}}
			\end{array}$
		\end{minipage}
		\caption{Oriented local loop vertex configurations. The corresponding weights reflect the connectivity type and the loop orientation.}
		\label{fig: oriented vertices}
	\end{figure}

	Alternatively, the oriented loop configurations can be constructed out of  eight  vertices with oriented arcs, see Fig. \ref{fig: oriented vertices}, generalizing the non-oriented ones from Fig. \ref{fig: loop-vertices}.  The vertices are placed onto lattice sites, so that the arrows  on the  bonds agree. 
	Within  this construction  the weights $q$ and $q^{-1}$ of a contractible oriented   loop reappear if the weight $q^{\frac{1}{4}}$ ($q^{-\frac{1}{4}}$) is ascribed to  every arc that makes the clockwise (counterclockwise)  ninety-degree turn, see Fig. \ref{fig: oriented vertices}. We also supply the weights with  the  dependence on an   auxiliary spectral  parameter  $z$ that  reflects  the arc connectivity within the vertex and should be set to $z=1$ in the end to return to the  $O(1)$ DLM.  
	
	Having fixed the contractible loop weights   we still have to take care of the weights of the defects, which were set equal to one in  $O(1)$ DLM defined in the beginning. In a finite lattice domain these weights depend on  the BCs under consideration. In the following we  consider the   $O(1)$ DLM  on the  infinite cylinder of odd circumference that excludes non-contractible loops closed around the cylinder because of the presence of  at least one infinite defect. To treat  the loop model on the infinite cylinder, we start with the model on the finite  torus and  send one of the periods to infinity. The model  on the torus still admits the non-contractible loops, which  become the infinite defects in the limit. The orientation along every such  directed loop is preserved when going around the torus so that the loop  has a unit weight when $z=1$. Respectively,  the weight  of an undirected  non-contractible loop, obtained from the summation over loop orientations,  is equal to two, unlike the unit weight in the initial definition accepted to keep the connection to the percolation. This  discrepancy, however, does not affect the infinite cylinder limit of the average values of quantities like the loop densities, to  which only the configurations with a single  defect bring a non-vanishing contribution. Alternatively, we could restrict the configuration set of the directed DLM to configurations with a fixed   direction of the non-contractible loops.  This amounts simply to   choosing  the sector with fixed difference between  numbers of up and down arrows in a horizontal row of vertical bonds.  
	
	Attaching the orientation of  arcs  to the bonds incident to a site independently of the arc connectivity gives the vertices of the  six-vertex model: the  four and the two of them  appear once and  twice respectively. For the latter, we sum up the weights of vertices of directed DLM with the same bond directions to obtain the weights of vertices of the corresponding asymmetric six-vertex model 
	\begin{eqnarray}
		&&a_1 = a_2  =  z, \notag \\
		\label{6V param}
		&&b_1 = b_2 = 1,\\
		&&c_1 = z q^{1/2}+q^{-1/2}, \notag\\ 
		&&c_2 =  q^{1/2}+zq^{-1/2}. \notag
	\end{eqnarray}
	So defined Boltzmann weights $a_1,a_2, b_1,b_2$  are real, when $z$ is real, while $c_1$ and $c_2$ are complex, if $q$ is complex. However, they enter all the weights of configurations on the cylinder or torus via  the product $c_1c_2$,  which is real in the case  $|q|=1$  under consideration.

	Let us consider the six-vertex model on a rectangular domain of the square lattice with periodic BCs in both directions, i.e. $\mathbb{Z}\slash L\mathbb{Z}\times \mathbb{Z}\slash H \mathbb{Z}$, where  $L,H\in \mathbb{N}$. We  denote its partition function, depending on the parameters $q,z$ by $Z^{(6V)}_{L,H}(q,z)$. According to the previous discussion, when $z=1$ it coincides with the partition function $Z_{L,H}^{(DLM)}(n)$ of the modification of  $O(n)$ DLM, where a  contractible  loop is assigned the weight $n$ given by  (\ref{eq: weight n}) and a non-contractible loop has the weight $2$.
	\begin{equation}
		Z^{(DLM)}_{L,H}(n)=Z^{(6V)}_{L,H}(q,1).
	\end{equation} 
	Then, we define the per-site free energy for the infinite cylinder as a limit
	\begin{equation}\label{eq: free energy}
		f_L(n)=\lim_{H\to \infty}  \frac{1}{LH} \log \left( Z^{(DLM)}_{L,H}(n) \right),
	\end{equation}
	whose derivative yields the average loop density in 
	\begin{equation}\label{eq: nu(L) via f(n)}
		\nu(L)= n \frac{d}{dn} f_L(n),
	\end{equation} 
	which we are going to  evaluate below for odd $L$ at $n=1$.

	To evaluate the partition function of the six-vertex model we define the row-to-row transfer matrix of the six-vertex model $	\mathbb{T}^{(6V)}_L(u):\mathcal{H}\to\mathcal{H}$ acting in the space $\mathcal{H}=(\mathbb{C}^2)^{\otimes L}$  that spans a
	spin basis  $\{\uparrow, \downarrow \}^{\otimes L}$, where the basis vectors  $\uparrow = (1,0)^T$ and $ \downarrow = (0,1)^T$ of every tensor component are associated with the up and down  arrows at the corresponding vertical bonds in a horizontal row. To this end, we first introduce linear operators $R_{ij}(u)$ 
	acting in an extended  space  $ \mathbb{C}^2 \otimes \mathcal{H}$. Their lower indices $i,j\in \{0,\dots, L\}$ refer to  the tensor components, among which  $L$ copies of $\mathbb{C}^2$  are  the factors of  $\mathcal{H}$ with indices $1,\dots,L$  associated with $L$ vertical bonds in the same horizontal row of the lattice  and the  one with index $0$ is an auxiliary space representing  the horizontal line. The operator   $R_{ij}(u)$  acts non-trivially in  the tensor factors with indices $i,j$ and as the identity operator in the other components.
	The non-trivial action in the  pair of spaces   recorded in the basis $\{\uparrow \uparrow, \uparrow  \downarrow,  \downarrow \uparrow,  \downarrow   \downarrow\}$ is given by the  $R$-matrix
	\begin{eqnarray}
		R(u) &=&   \begin{pmatrix}
			a(u) & 0 & 0 & 0\\
			0 & b(u) & c(u) & 0\\
			0 & c(u) & b(u) & 0\\
			0 & 0 & 0 &  a(u)
		\end{pmatrix} \notag\\&=& \begin{pmatrix}
			\frac{u-q}{1-qu} & 0 & 0 & 0\\
			0 & 1 & \sqrt{\frac{u}{q}} \frac{(1-q^2)}{(1-qu)} & 0\\
			0 & \sqrt{\frac{u}{q}}\frac{(1-q^2)}{(1-qu)} & 1 & 0\\
			0 & 0 & 0 &  \frac{u-q}{1-qu}
		\end{pmatrix} \label{eq: R-matrix}
	\end{eqnarray}
	with coefficients being the vertex weights \eqref{6V param}  reexpressed in terms of another spectral parameter $u$ obtained by the variable change 
	\begin{equation} \label{eq: z vs u}
		z = \frac{u-q}{1-qu} .
	\end{equation}
	Note that the matrix  $R(u)$ is an equivalent symmetric version of the  matrix defined by weights 
	(\ref{6V param}). Instead of two different weights $c_1$ and $c_2$, it contains their combination $c(u)=\sqrt{c_1c_2}$ on  which, as was mentioned before,  all the configuration weights depend.
	Algebraically, the two matrices are related by a simple conjugation in one of two tensor components where they act, which disappears  under the trace  operation used to define the transfer matrix below.      
	
	Then, for odd $L$ the transfer matrix is given by the trace over the auxiliary space  of the product of $R$-operators
	\begin{equation}\label{eq: 6V transfer matrix}
		\mathbb{T}^{(6V)}_L(u) = \mathrm{Tr_0} \left( R_{0L} (u) \dots R_{02}(u) R_{01} (u) \right)
	\end{equation}
	and the torus partition function is the trace over the other tensor components
	\begin{equation}\label{eq: Z^{(6V)}_{L,H}(q,z)=Tr}
		Z^{(6V)}_{L,H}(q,z)=\mathrm{Tr}  \left(\mathbb{T}_L^{(6V)} (u) \right)^{H}.  
	\end{equation}
	The choice (\ref{eq: z vs u}) of the spectral parameter  $u$ is useful for the $R$-operators to satisfy  the Yang-Baxter equation in the form
	\begin{equation}
		R_{12}(u/qv) R_{13}(u) R_{23}(v) = R_{23}(v) R_{13}(u) R_{12}(u/qv),
	\end{equation}
	This ensures  commutativity of the transfer matrices at different values of the spectral parameter  
	\begin{equation}
		\mathbb{T}^{(6V)}_{L}(u) \mathbb{T}^{(6V)}_{L}(v) = \mathbb{T}^{(6V)}_{L}(v) \mathbb{T}^{(6V)}_{L}(u),
	\end{equation}
	which is  the crucial fact for the exact solvability of the model. 
	
	We note that for even $L$,  the product under the trace in (\ref{eq: 6V transfer matrix}) is supplied with an additional factor
	$\Omega_0$ acting as $\mathrm{diag} (q^{-1}, q)$ in the auxiliary space and as the identity in the other tensor components. This factor  corrects the weight of the
	non-contractible loops on the cylinder of an even circumference. Since due to the defect  there are no non-contractible loops when $L$ is odd, this factor is not necessary (see more details in \cite{Zinn2009six, FZZ2006}). As we already mentioned, we  should have inserted similar operators  acting in $L$ tensor factors of $\mathcal{H}$ under the trace in (\ref{eq: Z^{(6V)}_{L,H}(q,z)=Tr}) to  correct the wrong weight of the non-contractible loops going once  around the torus  in the other direction and hence the weight of the defect on the infinite cylinder. However, this   does not affect the limiting expression (\ref{eq: free energy}) of the free energy anyway.
	
	In the infinite cylinder  limit (\ref{eq: free energy}) only the contribution of  the largest eigenvalue $\Lambda_{max}(u)$ of the transfer matrix $\mathbb{T}^{(6V)}_{L}(u) $ survives under the trace so that  the free energy is given by 
	\begin{equation}\label{eq: f(n) via Lambda}
		f_L(n) =   \frac{1}{L } \log \Lambda_{max} (1).
	\end{equation}
	
	Thus, the problem is reduced to  finding the largest eigenvalue  of the six-vertex transfer matrix. 
	Below, we employ the Bethe ansatz method to find the largest eigenvalue and its derivative at the stochastic point. Before going to certain formulas, let us discuss 
	the general structure of the spectrum of the transfer matrix and the place of the largest eigenvalue in it. 
	
	We first note that   the transfer matrix commutes with the operator  $S^z=\sum_{i=1}^L\sigma_i^z$, where $\sigma_i^z$ acts as the  Pauli matrix $\sigma^z=\mathrm{diag}(1,-1)$ in $i$-th component of $\mathcal{H}$ and as the identity in the others. In other words, the transfer matrix   preserves the number of up and down arrows in a transition between two subsequent horizontal rows of vertical bonds. Therefore,  the vector space  $\mathcal{H}$ can be decomposed to  a direct sum $\mathcal{H} = \oplus_M \mathcal{H}_M$ of invariant sub-spaces indexed by the total number $0\leq M \leq L$ of up arrows in the row.  Which subspace the largest eigenvalue belongs to is the first natural question to ask.
	
	Yang and Yang's seminal papers \cite{YY1, YY2} made  the first step in this direction. They studied the XXZ spin chain, whose Hamiltonian  shares its eigenvectors with the  six-vertex transfer matrix, and identified the  Bethe ansatz eigenstates, which    give the ground states  in   the invariant subspaces $\mathcal{H}_M$ with every $0\leq M \leq L$. 
	Based on these results, Lieb \cite{Lieb3} proved that in the disordered phase $|\Delta| \leq 1$ of the six-vertex model on the cylinder of circumference $L$, the largest eigenvalue of the transfer matrix  $\mathbb{T}_L^{(6V)} (z)$  belongs to an invariant subspace with the number of up arrows $M$ asymptotically equal to $L/2$ as $L\rightarrow \infty$. This equality is  also expected to be exact  for finite even  values of $L$.  For odd  $L$, the natural numbers closest to $L/2$ are both $(L \pm 1)/2$. Since the  subspaces $\mathcal{H}_M$ and $\mathcal{H}_{L-M}$ are related by the spin reversal symmetry  possessed by the transfer matrix, the  two-fold degenerate $\Lambda_{max}$ is expected  to  belong to both  the subspaces $\mathcal{H}_{(L+1)/2}$ and $\mathcal{H}_{(L-1)/2}$ in the odd case.
	Though there are still no rigorous proofs neither for  odd  nor for even  finite $L$, there is remarkable progress on the asymptotic behavior of the dominant eigenvalues, like, the bounds from above \cite{D-CKKMT}  and its ratios \cite{duminil2021discontinuity}.

\begin{figure}[h]
		\centering
		\setlength{\unitlength}{0.1\textwidth}
		\includegraphics[width=0.7\textwidth]{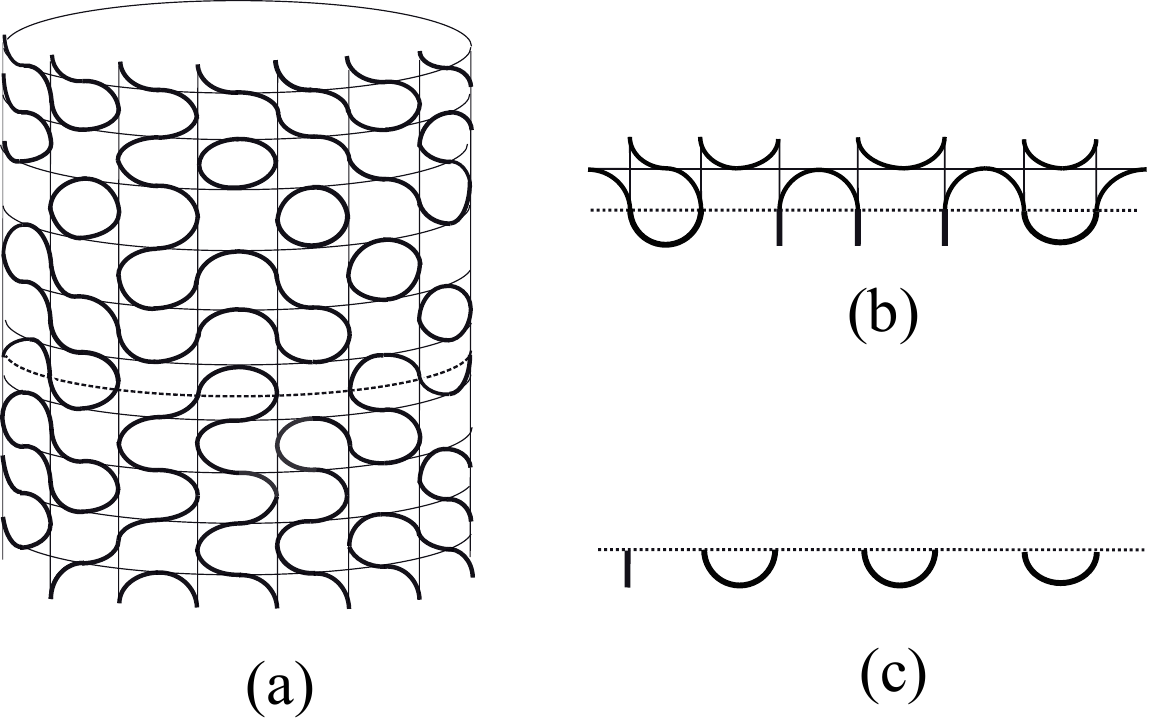}
		\caption{a) A bond cut of an infinite cylinder. b)The link pattern with three unpaired bonds corresponding to the lower half-cylinder with one row of vertices attached. c) The partial link pattern with one unpaired bond obtained after the concatenation.  }
		\label{fig: link pattern}
	\end{figure}

	Nevertheless,  in our particular case, the problem simplifies significantly  due to  the connection of the six vertex model  with the  $n=1$ case of  $O(n)$ DLM. The simplification comes from the fact that at the stochastic point the six-vertex transfer matrix  written in a certain basis  is reduced to the   $O(1)$ transfer matrix, which  in turn, up to a normalization factor of  $2^L$ has the form of the transition probability matrix of a Markov chain. 
	The states from the state space  of the chain  are  non-crossing  partial pairings of $L$ bonds, which 
	show  how vertical bonds in a horizontal row are connected by paths within the  half of the cylinder below, see Fig. \ref{fig: link pattern}.  They are pictured  by diagrams, which we refer to as partial link patterns. Thus, the vector space, in which the  $O(1)$ DLM transfer matrix acts,  spans the basis consisting of vectors indexed by the partial link patterns.
	
	The step of the chain  is a concatenation  of a random  row of vertices from Fig. \ref{fig: loop-vertices}, each  chosen independently  with  probability $1/2$,  on top of the semi-infinite cylinder. When a row of vertices is concatenated to a partial link pattern, it results in another partial link pattern with the same or smaller number of unpaired bonds, as shown in   Fig. \ref{fig: link pattern}(b)-(c).  Correspondingly,  the underlying vector space  has a nested structure with respect to the action of the transfer matrix.

	Being  irreducible and aperiodic the Markov transition matrix has a unique  stationary state in the subspace with a minimal number of unpaired bonds, which is zero when $L$ is even and is one when $L$ is odd. It   corresponds to  the non-degenerate  Perron-Frobenius  eigenvalue equal to one. Thus,  the largest eigenvalue of the   $O(1)$ DLM transfer matrix is $2^L$, which is the number of different   ways one can  construct a  row of  $L$ vertices from those in  Fig. \ref{fig: loop-vertices}.     
	Being non-degenerate, the largest eigenvalue is also analytic in $n$ in a vicinity of $n=1$. 
	
\begin{figure}[H] 
		\begin{minipage}{1\linewidth}
			\centering
			\inlinegraphicsformula{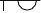}$ \ \ \ \rightarrow \ \ \ q^{\frac{1}{2}}$ \inlinegraphicsformula{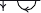} $+ \ q^{-\frac{1}{2}}$ \inlinegraphicsformula{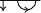} $+\  q^{\frac{1}{2}}$ \inlinegraphicsformula{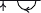} $+\ q^{-\frac{1}{2}}$ \inlinegraphicsformula{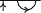} \\ 
			\vspace{0.3cm}
			$ 
			\rightarrow \ \ \ q^{\frac{1}{2}}$ \inlinegraphicsformula{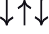} $ 
			+ \ q^{-\frac{1}{2}}$ \inlinegraphicsformula{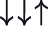} $+ \  q^{\frac{1}{2}}$  \inlinegraphicsformula{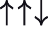} $ 
			+ \ q^{-\frac{1}{2}}$ \inlinegraphicsformula{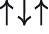}
		\end{minipage}
		\caption{An example of the correspondence between a basis element of DLM transfer matrix $\mathbb{T}_L^{(DLM)}$, namely partial link pattern, and a linear combination of spin basis elements. }
		\label{fig:from link to spin}
	\end{figure}

	To make sure that this is the same largest eigenvalue as the one of the corresponding six-vertex model, one has to identify the space that spans the  partial link basis with the subspace of the space $\mathcal{H}$ associated with the six vertex model. 
	This is done as follows. A basis element corresponding to a partial link pattern is represented as a linear combination of spin basis elements. They are  obtained by giving   all possible orientations to  half-loops and defects, supplying  every clockwise (counterclockwise) half-loop and a directed defect with the weight $q^{ 1/2}$ ($q^{-1/2}$) and   the unit weight respectively. Then, the orientation is transferred onto the bonds, see e.g. Fig.\ref{fig:from link to spin}. 
	It is then obvious  that the   sector with minimal number of unpaired vertical bonds belongs to the subspace  $\mathcal{H}_{L/2}$, when $L$ is even and to the union of $H_{(L+1)/2}$ and $H_{(L-1)/2},$ when $L$ is odd. 
	Since both are the invariant subspaces related by the spin reversal symmetry respected by the six-vertex transfer matrix, it is enough to limit the consideration to one of them. This is equivalent to considering undirected loops with a single directed defect, which, in particular, will have the correct unit weight, instead of  the weight $2$ of the undirected one.  
	See more details on the spin-link correspondence  in \cite{MitraNienhuis, FZZ2006, Zinn2009six, RSZ2007,morin2013homomorphism}.

	\section{Bethe ansatz}

	The eigenvectors of the transfer matrix $\mathbb{T}^{(6V)}_{L}(u)$ belonging to  every subspace $\mathcal{H}_M$ are constructed with the standard machinery of the algebraic   Bethe ansatz \cite{faddeev1996algebraic}. They are obtained  from the upper  vacuum vector with  all spins up $(\uparrow)^{\otimes L}$ by the action of $M$ lowering  operators, each depending on one spectral parameter $u_j$. The constructed vectors correspond to  the eigenvalue
	\begin{equation}\label{eq: Lambda_M}
		\Lambda_M(u) =  a(u) ^L  \prod_{j=1}^M \frac{a(u_j/qu) }{ b(u_j/qu )} +  b(u)^L \prod_{j=1}^M \frac{a(u/qu_j) }{ b(u/q u_j )},
	\end{equation}
	when the parameters $u_1, \dots, u_M$ solve the system of the Bethe equations
	\begin{equation}
		\frac{a(u_j)^L}{b(u_j)^L} = - \prod_{k=1, \neq j}^M \frac{a(u_j/qu_k)}{b(u_k/qu_j)} \frac{b(u_k/qu_j)}{a(u_j/qu_k)} , \quad j = 1, \dots M.
	\end{equation}
	After substitution of the  vertex weights \eqref{eq: R-matrix} the eigenvalues of $\mathbb{T}^{(6V)}_L(u)$ become
	\begin{equation} \label{BA-eigenvalue}
		%	\Lambda_M(x) = \left(\frac{1-q^2x^2}{q(x^2-1)} \right)^L \prod_{j=1}^M \frac{x^2-q^2x_j^2}{q(x_j^2-x^2)} + \prod_{j=1}^M  \frac{x_j^2-q^2x^2}{q(x^2-x_j^2)} 
		\Lambda_M(u) = \left( \frac{u-q}{1-q u} \right)^L  \prod_{j=1}^M \frac{u_j - q^2 u}{q(u-u_j)} + \prod_{j=1}^M  \frac{u - q^2 u_j}{q(u_j-u)}, \end{equation}
	where $u_1, \dots, u_M$ is a solution of the system of  Bethe equations
	\begin{equation} \label{eq: BAE}
		%	\left( \frac{1-q^2x_j^2}{q(x_j^2-1)} \right)^L = (-1)^{M-1} \prod_{k=1}^M \frac{x_k^2 - q^2x_j^2 }{x_j^2 - q^2x_k^2}, \quad j = 1, \dots M.
		%
		\left( \frac{u_j-q}{1-q u_j} \right)^L = (-)^{M-1} \prod_{k=1}^M \frac{u_j - q^2u_k }{u_k - q^2u_j}, \quad j = 1, \dots M.
	\end{equation}
	In the particular case of $L=1$, the products in r.h.s. of  (\ref{eq: Lambda_M}) are empty and the only eigenvalue of the transfer matrix is $\Lambda_1= a(u)^L + b(u)^L$. It does not depend on $q$ when  $u = 1$. Thus, we obtain $\nu(1) = 0$.
	
	\subsection{Functional Bethe ansatz}
	
	The eigenvalue  \eqref{BA-eigenvalue} can be reformulated in terms of a Q-polynomial of degree $M$ having zeroes at  Bethe roots 
	\begin{equation}
		Q(u) = \prod_{k=1}^M (u - u_k).
	\end{equation}
	Then, the system of $M$ Bethe equations \eqref{eq: BAE} can be reformulated as one functional equation for polynomials $T(u)$
	and $Q(u)$
	\begin{equation} \label{eq_TQ} 
		T(u)Q(u) =  \phi(q^{-1}u) Q(q^2u) + \phi(q u) Q(q^{-2}u) (-q)^{2M-L},  
	\end{equation}
	where
	\begin{equation} \label{eq: lambda_via_T}
		T(u) = \Lambda(u) \phi(qu) (-q)^{M-L}
	\end{equation}
	and 
	\begin{equation}
		\phi(u) = (1-u)^L.
	\end{equation} 
	A conjugate problem arises when one constructs the eigenvectors in $\mathcal{H}_M$ with the algebraic Bethe ansatz starting from the lower vacuum vector $(\downarrow)^{\otimes L}$ and acting at it by  $L-M$ raising   operators. In general, this procedure is equivalent to the one described above  applied to another asymmetric six-vertex model, obtained from the first one by an exchange of  the  weights $(a_1,b_1,c_1)$ and $(a_2,b_2,c_2)$ in the subspace $\mathcal{H}_{L-M}$.  In our case of the symmetric
	$R$-matrix $R(u)$, for which the triples of parameters coincide,  it is enough to use already obtained  T-Q equation written for the subspace $\mathcal{H}_{L-M}$  requiring that the eigenvalue $\Lambda(u)$ is still the same.   
	
	Thus, one  obtains another functional equation   
	\begin{equation} \label{eq_TP}
		T(u)P(u) = (-q)^{2M-L}  \phi(q^{-1}u)P(q^2 u) +  \phi(qu) P(q^{-2}u).
	\end{equation}
	Multiplying equations \eqref{eq_TQ} and \eqref{eq_TP} by $P(u)$ and $Q(u)$, respectively, subtracting one from the other and analyzing the structure of poles, we arrive at the quantum  Wronskian relation between $P(u)$ and $Q(u)$
	\begin{equation} \label{phi(u) from TQ}
		\phi(u) =  \frac{ Q(qu)P(q^{-1}u) -  Q(q^{-1}u)P(qu)(-q)^{2M-L}}{ (-q)^{2M-L} -1}.
	\end{equation}
	Substituting this expression into \eqref{eq_TQ} or \eqref{eq_TP} we have
	\begin{equation} \label{eq: T(u) from TQ}
		T(u) =  \frac{ Q(q^2u)P(q^{-2}u) -  Q(q^{-2}u)P(q^2u)(-q)^{4M-2L}}{(-q)^{2M-L} -1}.
	\end{equation}
	We note that $T$-$Q$ and $T$-$P$ equations are  the ones  obtained in \cite{Str2001importance} for $\Delta = -\frac{1}{2}$ XXZ model written in  different  parametrization  $q= -e^{i \eta}, u = e^{2 i \tilde{u}}$.
	
	\subsection{FSZ solution to T-Q and T-P equations}
	Now we reproduce the solution of  $T$-$Q$ and $T$-$P$ obtained by Fridkin, Strognov, Zagier \cite{FSZ2000even, FSZ2001even} for even $L$ and Stroganov \cite{ Str2001importance} for odd $L$  
	\begin{equation}
		M = N = \frac{L-1}{2},
	\end{equation}
	corresponding to the dominant eigenvalue of the six-vertex transfer matrix at the stochastic point \eqref{stoch_point}.  Below we always mean $q= e^{\frac{\mathrm{i}\pi}{3}}$ and use $q^3 = -1,$ where necessary.
	
	Let us define functions $Q_k = Q(q^{2k} u), \phi_k = \phi(q^{2k-1} u), T_k = T(q^{2k} u)$ with integer $k$. The T-Q equation \eqref{eq_TQ} is equivalent to the homogeneous system of three linear in $Q_k$ equations
	\begin{equation}
		T_k Q_k = \phi_k Q_{k+1} + \phi_{k+1} Q_{k-1} q^2, \quad k=0,1,2.
	\end{equation}
	According to Fridkin, Strognov, and Zagier, for the ground state at the stochastic point, the rank of the matrix for this system is one, which immediately gives 
	\begin{equation} \label{T(u) at st point}
		T_k = q \phi_{k-1} \Longleftrightarrow T(u) = q(1+u)^L.
	\end{equation}
	We define the polynomial 
	\begin{equation}\label{eq: f_Q}
		f_Q(u) = (1+u)^L Q(u)
	\end{equation}
	of degree at most $3N+1.$ The T-Q equation is equivalent to the equation
	\begin{equation} \label{eq_f(u)}
		f_Q(u) + q^2 f_Q(q^2 u) + q^4 f_Q(q^4 u) = 0
	\end{equation}
	that causes some coefficients of  
	\begin{equation}
		f_Q(u) = \sum_{k=0}^{3N+1} f_k u^k.
	\end{equation}
	to vanish,
	\begin{equation}
		f_{3m+2} = 0, \quad m = 0, \dots, N-1.
	\end{equation}
	The requirement of having a zero at $u = -1$ of order $2N+1$ is a set of additional $2N+1$ equations
	that fixes the other coefficients of $f_Q(u)$ yielding 
	\begin{eqnarray}
		f_Q(u) &=&\nonumber
		(-1)^N u \frac{\Gamma(\frac{2}{3})\Gamma(N+\frac{4}{3})}{\Gamma(\frac{4}{3}) \Gamma(N+\frac{2}{3})} \  _2F_1(-N, \frac{1}{3}-N, \frac{4}{3}, -u^3) \\ \label{f_Q}
		&+& (-1)^N \ _2F_1(-N, -\frac{1}{3}-N, \frac{2}{3}, -u^3). 
	\end{eqnarray}
	Similarly, the T-P equation written for 
	\begin{equation}\label{eq: f_P}
		f_P(u) = (1+u)^L P(u)
	\end{equation}
	is solved by 
	%is equivalent to another cyclic relation
	%\begin{equation}
	%	f_{P}(u) + q^{-2}f_{P}(q^2u)+q^{-4}f_{P}(q^{4}u) = 0
	%\end{equation}
	%on the polynomial $f_{P} = (1+u)^L P(u)$. With analogous techniques, we find
	\begin{eqnarray} \nonumber
		f_P(u) &=& (-1)^N u^2 \frac{\Gamma(\frac{1}{3})\Gamma(N+
			\frac{5}{3})}{\Gamma(\frac{5}{3})\Gamma(N+\frac{1}{3})} \ _2F_1(-N, \frac{2}{3}-N, \frac{5}{3}, -u^3) \\ \label{f_p}
		&-& (-1)^N \ _2F_1(-N, -\frac{2}{3}-N, \frac{1}{3}, -u^3).
	\end{eqnarray}   
	See \cite{FSZ2000even, FSZ2001even, Str2001importance,Pov2021}  for details.
	
	One can easily check that $f_Q(u)$ and $f_P(u)$ satisfy to differential equations 
	\begin{eqnarray*}
		(1+u^3)  f''_Q(u) -6Nu^2 f'_Q(u) + 3N(3N+1)u f_Q(u)&  =& 0,	\\
		(1+u^3) u f''_P(u) -\left(u^3 (6N+1)+1\right) f'_P(u) + 3N(3N+2)u^2 f_P(u) & = & 0,
	\end{eqnarray*}
where we use the notation $f'(u) = \partial_u f(u)$ for the defivative of finction $f(u)$.

	\subsection{Calculation of derivative}
	In the next section, we find the density of loops calculating the derivative of the largest eigenvalue that we express in terms of the logarithmic derivative of $T(1)$ by the parameter $q$ using (\ref{eq: weight n}, \ref{eq: free energy}, \ref{eq: nu(L) via f(n)}, \ref{eq: f(n) via Lambda}, \ref{eq: lambda_via_T}) at the stochastic point \eqref{stoch_point}. We simplify the expressions with $q^{3} = -1$, $q^2 = q-1$, etc., where necessary.
	\begin{equation} \label{nu_via_T}
		\nu (L)= \frac{1}{2}+\frac{1}{2Lq(1+q)} + \frac{1}{L(1+q)} \left[\frac{d }{dq} \log T(1)\right] \Big|_{q=e^{\frac{\mathrm{i}\pi}{3}}}
	\end{equation}
	To differentiate $T(1)$ in $q$, we use its expression given by RHS of \eqref{eq: T(u) from TQ} and obtain
	\begin{equation} \label{eq: dT/dq}
		\frac{d T(1)}{dq} \Big|_{q=e^{\frac{\mathrm{i}\pi}{3}}} = 2A -\frac{2 Q(q^{-2})P(q^{2})}{q^2-1} + \frac{ q T(1)}{q^2-1} + B.
	\end{equation}
	where we introduce notation
	\begin{eqnarray*}
		A =  \left(q^2-1\right)^{-1} \Big( q Q'(q^2)P(q^{-2})+Q(q^2)P'(q^{-2}) \\
	&&\!\!\!\!\!\!\!\!\!\!\!\!\!\!\!\!\!\!\!\!\!\!\!\!\!\!\!\!\!\!\!\!\!\!\!\!\!\!\!\!\!\!\!\!\!\!\!\! \!\!\!\!\!\!\!\!\!\!\!\!\!\!\!\!\!\!\!\!\!\!\!\!\!\!\!\!\!\!\!\!\!\!\!\!\!\!\!\!\!\!\!\!\!\!\!\!\!\!\!\!\!\!\!\!\!\ + q Q'(q^{-2})P(q^2) +q^{2}Q(q^{-2})P'(q^2 ) \Big).
	\end{eqnarray*}
	The first three summands of \eqref{eq: dT/dq} come from the explicit dependence of r.h.s.  of \eqref{eq: T(u) from TQ} on $q$.
	The forth term denoted by  the letter $B$ is the contribution from unknown dependence of the Bethe roots on the parameter $q$. 
	
	The same term  $B$ appears when we evaluate the derivative of the Wronskian relation \eqref{phi(u) from TQ} 
	at the special value of the spectral parameter $u =-1$ with respect to $q$ which should then be set to $q=e^{i\pi/3}$.
	\begin{equation} \label{eq:dphi/dq}
		\frac{d \phi(-1) }{dq} \Big|_{q=e^{\frac{\mathrm{i}\pi}{3}}} = \frac{1}{q} \left(-A + \frac{q^2 Q(q^2)P(q^{-2})}{q^2-1} + B + \frac{q^2 \phi(-1)}{q^2-1} \right) = 0
	\end{equation}
	Expressing $B$ from the equation \eqref{eq:dphi/dq}  and using (\ref{T(u) at st point}) we arrive at
	\begin{equation}\label{eq: dln T/dq}
		\frac{d \ \log T(1)}{dq} \Big|_{q=e^{\frac{\mathrm{i}\pi}{3}}} = \frac{1}{q\, 2^L}\left( 3A -\frac{q^2 Q(q^2)P(q^{-2})+2Q(q^{-2})P(q^2)}{q^2-1}\right).
	\end{equation}
	Our next step is to express the values of polynomials $Q(q^{\pm^2})$, $P(q^{\pm^2})$, $Q'(q^{\pm^2})$, $P'(q^{\pm^2})$. Using (\ref{eq: f_Q}) and (\ref{eq: f_P})  we rewrite (\ref{eq: dln T/dq}) in the form
	\begin{eqnarray} \nonumber
		\frac{d \log T(1) }{dq} \Big|_{q=e^{\frac{\mathrm{i}\pi}{3}}} &=& \frac{1}{q 2^L(q^2-1)} 
		\Big[3 \Big( q f'_Q(q^2)f_P(q^{-2})+f_Q(q^2)f'_P(q^{-2}) \\ \nonumber
		&+&q f'_Q(q^{-2})f_P(q^{2})
		+q^{2} f_Q(q^{-2})f'_P(q^{2}) \\ \nonumber
		&-&L(1+q) \left(f_Q(q^2) f_P(q^{-2})  + q f_Q(q^{-2}) f_P(q^2)\right) \Big) \\ 
		&-&\Big(q^2 f_Q(q^2) f_P(q^{-2}) + 2f_Q(q^{-2})f_P(q^{2})\Big)
		\Big].   \label{dT/dq}
	\end{eqnarray}
	The last and the most technical part is the calculation of $f_{P}(q^{\pm 2})$, $f_{Q}(q^{\pm 2})$, $f'_{P}(q^{\pm 2})$, $f'_{Q}(q^{\pm 2})$ and simplification of \eqref{nu_via_T}.  We divide it into three parts. First, one calculates the derivatives of $f_Q(u)$ and $f_P(u)$ with the differentiation formula 
	\begin{equation}
		\frac{\partial}{\partial z} \   _2F_1(a,b,c,z) = \frac{a b}{c} \  _2F_1(a+1,b+1,c+1,z)
	\end{equation}
	applied to \eqref{f_p} and \eqref{f_Q}.
	Second, one evaluates all four polynomials at $q^{\pm 2}$  with generalization of  Kummer's theorem 
	\begin{eqnarray}
		&&\!\!\!\!\!\!\!\!\!\!\!\!\!\!\!\!\!\!\!\!\!\!\!\!	\ _2F_1(a,b, 1+a-b+n, -1)\\& =& \frac{ \Gamma(1+a-b+n) \Gamma(1-b)}{2 \Gamma(a) \Gamma(1-b+n)} \sum_{k=0}^n (-)^k \binom{n}{k} \frac{\Gamma(\frac{a}{2}+\frac{k}{2})}{\Gamma(\frac{a}{2}+\frac{k}{2}-b+1)},\notag
	\end{eqnarray}
	proved in \cite{ChK2007} for every integer $n$. We meet only $n=0$ and $n=1$ appearing after the differentiation. The final expressions in terms of gamma functions have the form
	\begin{equation}
		f_Q(q^{ \pm2}) = (-)^N g(q^{ \pm2}, 1), \qquad  f_P(q^{ \pm2}) = (-)^Ng(q^{ \pm2}, 2)
	\end{equation}
	\begin{equation}
		f'_Q(q^{ \pm2}) = (-)^N h(q^{ \pm2}, 1), \qquad  f'_P(q^{ \pm2}) = (-)^Nh(q^{ \pm2}, 2)
	\end{equation}
	where 
	\begin{eqnarray}\nonumber
		g(x, k) =   \frac{x^{\pm k}
			\Gamma(\frac{3-k}{3}) 
			\Gamma(\frac{3+k}{3}+N) \Gamma(\frac{k}{6}-\frac{N}{2}) } {2\Gamma(\frac{3-k}{3}+N) \Gamma(\frac{k}{3}-N) \Gamma(\frac{6+k}{6}+\frac{N}{2})} 
		+ \frac{ (-)^k 
			\Gamma(\frac{3-k}{3}) \Gamma(-\frac{k}{6}-\frac{N}{2})} {2 \Gamma(-\frac{k}{3}-N) \Gamma(\frac{6-k}{6}+\frac{N}{2})},
	\end{eqnarray}
	%\begin{eqnarray} \nonumber
	%f_P(q^{\pm2}) = (-)^N  \frac{q^{\mp 2}
		%	\Gamma(\frac{1}{3}) 
		%	\Gamma(\frac{5}{3}+N) \Gamma(\frac{1}{3}-\frac{N}{2})} {2\Gamma(\frac{1}{3}+N) \Gamma(\frac{2}{3}-N) \Gamma(\frac{4}{3}+\frac{N}{2})} 
	%-(-)^N \frac{
		%	\Gamma(\frac{1}{3}) \Gamma(-\frac{1}{3}-\frac{N}{2})} {2 \Gamma(-\frac{2}{3}-N) \Gamma(\frac{2}{3}+\frac{N}{2})},
	%\end{eqnarray}

	\begin{eqnarray*}
		h(x, k ) &=&\!\!\!  x^{k} \frac{
			\Gamma(\frac{3-k}{3}) 
			\Gamma(\frac{3+k}{3}+N) \Gamma(\frac{k}{6}-\frac{N}{2})} {2\Gamma(\frac{3-k}{3}+N) \Gamma(\frac{k}{3}-N) \Gamma(\frac{6+k}{6}+\frac{N}{2})}\\ 
		&-& \!\!\! x^{k} \frac{ (3N-1-k)
			\Gamma(\frac{3-k}{3}) \Gamma(N + \frac{3+k}{3})}{2 \Gamma(\frac{3-k}{3}+N) \Gamma(\frac{3+k}{3}-N)} 
		\left(
		\frac{\Gamma (\frac{3+k}{6}-\frac{N}{2})}{\Gamma (\frac{3+k}{6}+\frac{N}{2})} -\frac{\Gamma(\frac{6+k}{6}-\frac{N}{2}) }{\Gamma (\frac{6+k}{6}+\frac{N}{2})}\right) \\
		&-& \!\!\! \frac{ (-)^{k-1} 3 x^{\pm 2} (3N+k)
			\Gamma(\frac{6-k}{3})}{4 \Gamma(\frac{3-k}{3}-N)} 
		\left(
		\frac{\Gamma (\frac{3-k}{6}-\frac{N}{2})}{\Gamma (\frac{3-k}{6}+\frac{N}{2})} -\frac{\Gamma(\frac{6-k}{6}-\frac{N}{2}) }{\Gamma (\frac{6-k}{6}+\frac{N}{2})}\right).
	\end{eqnarray*}
	%\begin{eqnarray*}
	%f'_P(q^{ \pm 2}) &=&  q^{ \pm 2} \frac{
		%		\Gamma(\frac{1}{3}) 
		%		\Gamma(\frac{5}{3}+N) \Gamma(\frac{1}{3}-\frac{N}{2})} {\Gamma(\frac{1}{3}+N) \Gamma(\frac{2}{3}-N) \Gamma(\frac{4}{3}+\frac{N}{2})}\\ 
	%	&-&  q^{ \pm 2} \frac{ (3N-2)
		%		\Gamma(\frac{1}{3}) \Gamma(N + \frac{5}{3})}{2 \Gamma(\frac{1}{3}+N) \Gamma(\frac{5}{3}-N)} 
	%	\left(
	%	\frac{\Gamma (\frac{5}{6}-\frac{N}{2})}{\Gamma (\frac{5}{6}+\frac{N}{2})} -\frac{\Gamma(\frac{4}{3}-\frac{N}{2}) }{\Gamma (\frac{4}{3}+\frac{N}{2})}\right) \\
	%	&+&    \frac{ 3q^{\pm 4} (3N+2)
		%		\Gamma(\frac{4}{3})}{2 \Gamma(\frac{1}{3}-N)} 
	%	\left(
	%	\frac{\Gamma (\frac{1}{6}-\frac{N}{2})}{\Gamma (\frac{1}{6}+\frac{N}{2})} -\frac{\Gamma(\frac{2}{3}-\frac{N}{2}) }{\Gamma (\frac{2}{3}+\frac{N}{2})}\right).
	%\end{eqnarray*}
	Finally, we substitute these expressions into \eqref{dT/dq} and \eqref{nu_via_T} to obtain  rational combinations of gamma functions and  use the recurrent  property of   gamma functions
\begin{equation}
\Gamma(z+1) = z \Gamma(z),
\end{equation}	
the reflection identity
\begin{equation}
		\Gamma(z) \Gamma(1-z) = \frac{\pi}{ \sin (\pi z) },
\end{equation}
 and  Legendre duplication formula
	\begin{equation}\
\Gamma(z)\Gamma(z+\frac{1}{2}) = 2^{1-2z} \sqrt{\pi} \Gamma(2z)
	\end{equation}
	to express the gamma functions at different values of of argument through each other and simplify it up to the form \eqref{eq: density_result_in_Gamma}.
	
\section*{Acknowledgement}
We thank the anonymous referee for providing insightful comments that significantly enhanced the quality of the work. We also extend our gratitude to Robert M Ziff and Robert Shrock for valuable discussions and suggestions.	

\printbibliography
	
\end{document}